\documentclass[a4paper,aps,prd,10pt,preprintnumbers,twocolumn,superscriptaddress,nofootinbib,amsmath,amssymb]{revtex4-1}
\usepackage{graphicx}
\usepackage{hyperref}
\usepackage{subfigure}
\usepackage{multirow}
\usepackage[toc,page]{appendix}
\usepackage{ulem}
\usepackage{cases}
\usepackage{cmap}
\usepackage{color}

\DeclareMathAlphabet{\pazocal}{OMS}{zplm}{m}{n}

\def\imo{i}
\def\re#1{{\rm Re}(#1)}
\def\im#1{{\rm Im}(#1)}
\def\K{{\cal K}}

\def\nq{\hspace*{-1em}}

\def\nhq{\hspace*{-0.5em}}

\def\cm{\hspace*{1cm}}

\def\Jl#1#2{#1 {\bf #2},\ }

\def\ApJ#1 {\Jl{Astroph. J.}{#1}}
\def\CQG#1 {\Jl{Class. Quantum Grav.}{#1}}
\def\DAN#1 {\Jl{Dokl. AN SSSR}{#1}}
\def\GC#1 {\Jl{Grav. \& Cosmol.}{#1}}
\def\GRG#1 {\Jl{Gen. Rel. Grav.}{#1}}
\def\JETF#1 {\Jl{Zh. Eksp. Teor. Fiz.}{#1}}
\def\JETP#1 {\Jl{Sov. Phys. JETP}{#1}}
\def\JHEP#1 {\Jl{JHEP}{#1}}
\def\JMP#1 {\Jl{J. Math. Phys.}{#1}}
\def\NPB#1 {\Jl{Nucl. Phys.}{B\ #1}}
\def\NP#1 {\Jl{Nucl. Phys.}{#1}}
\def\PLA#1 {\Jl{Phys. Lett.}{#1A}}
\def\PLB#1 {\Jl{Phys. Lett.}{#1B}}
\def\PRD#1 {\Jl{Phys. Rev.}{D\ #1}}
\def\PRL#1 {\Jl{Phys. Rev. Lett.}{#1}}

\def\lal{&&\nq {}}
\def\beq{\begin{equation}}
\def\eeq{\end{equation}}
\def\bear{\begin{eqnarray}}
\def\bearr{\begin{eqnarray} \lal}
\def\ear{\end{eqnarray}}
\def\eq{Eq.\,}
\def\eqs{Eqs.\,}

\def\nnn{\nonumber\\ \lal }

\def\yyy{\\[5pt] \lal }

\def\dst{\displaystyle}
\def\tst{\textstyle}
\def\fracd#1#2{{\dst\frac{#1}{#2}}}
\def\fract#1#2{{\tst\frac{#1}{#2}}}
\def\Half{{\fracd{1}{2}}}
\def\half{{\fract{1}{2}}}

\def\d{\partial}

\def\wh{wormhole}
\def\whs{wormholes}

\def\asflat{asymptotically flat}

\def\Sch{Schwarzschild}

\def\thd{\fract 13}

\def\mn{_{\mu\nu}}

\def\mN{_\mu^\nu}

\def\schd{\biggl(1-\frac{2 M}{r}\biggr)}

\def\Z{{\mathbb Z}}
\begin{document}

\title{Echoes in brane worlds: ringing at a black hole--wormhole transition}

\author{Kirill A. Bronnikov}
\affiliation{VNIIMS, Ozyornaya ul. 46, Moscow 119361, Russia}
\affiliation{Inst. of Gravitation and Cosmology, RUDN University,
              ul. Miklukho-Maklaya 6, Moscow 117198, Russia}
\affiliation{National Research Nuclear University ``MEPhI'', Kashirskoe sh. 31, Moscow 115409, Russia}
\email{kb20@yandex.ru}

\author{Roman A. Konoplya}\email{roman.konoplya@gmail.com}
\affiliation{Institute of Physics and Research Centre of Theoretical Physics and Astrophysics,
   Faculty of Philosophy and Science, Silesian University in Opava, CZ-746 01 Opava, Czech Republic}
\affiliation{Inst. of Gravitation and Cosmology, RUDN University,
              ul. Miklukho-Maklaya 6, Moscow 117198, Russia}
\begin{abstract}
  Echoes are known as modifications of the usual quasinormal ringing of a black hole at late times
  because of the deviation of space-time from the initial black-hole geometry in a small region near
  its event horizon. We consider a class of brane-world model solutions of the
  Shiromizu-Maeda-Sasaki equations, which describe both black holes and wormholes and
  interpolate between them via a continuous parameter.
  In this way the brane-world scenario provides a natural model for wormholes mimicking the black
  hole behavior if the continuous parameter is chosen near the threshold with a black-hole solution.
  We show that in the vicinity of this threshold interpolating between black holes and wormholes,
  quasinormal ringing of the wormholes at the initial stage is indistinguishable from that of
  the black holes with nearby values of the above parameter,
  but at later times the signal is modified by intensive echoes. We notice that the black-hole
  mimickers that are wormholes near the threshold have the largest quality factor and are therefore
  the best oscillators among the considered examples.
\end{abstract}
\maketitle

\section{Introduction}

  Recent observations of black holes (BHs) and stars in the electromagnetic and gravitational
  channels \cite{Abbott:2016blz,TheLIGOScientific:2016src,Goddi:2017pfy,Akiyama:2019cqa}
  allow us to test the regime of strong gravity. Nevertheless, nowadays there is still large uncertainty
  in measuring the parameters of compact objects, such as their mass and angular momentum,
  which leaves considerable freedom for various interpretations of the geometries either towards
  BHs in modified theories of gravity or even in favor of such exotic objects as stable
  Schwarzschild stars or wormholes
  \cite{Damour:2007ap,Konoplya:2016pmh,Konoplya:2019nzp,Camilo:2018goy,Wei:2018aft,
  Berti:2018vdi,Cardoso:2016rao}.

 A wormhole geometry can be designed ad hoc in such a way that it would mimic the behavior of
 a BH in any astrophysically relevant processes (except for the Hawking radiation which is
 unlikely to be observed for large BHs) \cite{Damour:2007ap}. This is possible because
 the geometry of such a wormhole can be indistinguishable from that of
 the BH in the whole space outside a tiny region near the wormhole throat. At intermediately
 late times, classical radiation in the vicinity of a BH or wormhole must be dominated by
 the characteristic damped oscillations, quasinormal modes, which have been extensively studied
\cite{Konoplya:2011qq,Kokkotas:1999bd,Berti:2009kk}
 and were recently observed in, apparently, mergers of two BHs
\cite{Abbott:2016blz,TheLIGOScientific:2016src}.

  While quasinormal modes of wormholes mimicking BHs are almost the same, the
  signal is modified at later times by echoes, what was first observed in  \cite{Cardoso:2016rao}
  and further studied in a large number of papers (see, for example,
\cite{Cardoso:2016oxy,Cardoso:2017cqb,Tsang:2018uie,Konoplya:2018yrp,Barausse:2014tra,
  Nakano:2017fvh,Testa:2018bzd,Mirbabayi:2018mdm,Wang:2018mlp,Bueno:2017hyj,Maselli:2017tfq,Li:2019kwa,
  Wang:2019szm,Cardoso:2019apo}  and references therein).
  In \cite{Cardoso:2016rao,Cardoso:2016oxy} the echoes produced by various toy models of
  thin-shell wormholes were studied. A similar picture of echoes at the beginning of
  the threshold between a regular BH and a wormhole has been found in
  \cite{Churilova:2019cyt} for the metric suggested in \cite{Simpson:2018tsi}. There, a
  continuous parameter of the metric allowed for interpolation between a regular BH,
  a one-way wormhole with an extremal null throat (the so called black bounce) and a traversable
  wormhole. However, the above wormhole metric is not a solution of any field equations and was
  simply designed ad hoc in the same manner as the Damour-Solodukhin wormhole
  \cite{Damour:2007ap} whose echoes were studied in \cite{Cardoso:2016rao}.

  A few examples of families of exact solutions containing both BHs and wormholes have been
  found with phantom scalar fields as sources of gravity (see, e.g.,  \cite{pha1, pha2, pha3, pha4}),
  however, such solutions have been shown to be generically unstable under radial perturbations
  \cite{pha4, stab1, stab2, Bronnikov:2012ch}.

 Here we will also consider various geometries interpolating between BHs and traversable
  wormholes via a continuous parameter, but which, unlike the previous models, are exact solutions
  of the Shiromizu-Maeda-Sasaki equations \cite{maeda99}, describing the on-brane
  gravitational field in the second Randall-Sundrum brane-world scenario (RS2) \cite{ransum2}.
  The exact solutions
  to be studied concerning the possible echoes and quasinormal modes were obtained in
  \cite{bwh1, Bronnikov:2003gx}, and quasinormal modes in the frequency and time domains were
  studied in \cite{Abdalla:2006qj}, but only for the range of parameters representing BHs.
  Thus the effect of echoes which should take place for wormholes in the parameter range near
  the threshold with BHs was missed in \cite{Abdalla:2006qj}.

  Having all the above motivations in mind, we would like to consider the quasinormal modes and
  ringing profiles for a few examples of the BH/wormhole solutions of the
  Shiromizu-Maeda-Sasaki equations in order to understand the possible general imprints of this
  extra-dimensional scenario on the ringing profile of BHs and wormholes, and especially
  at the transition between them. We will show that once the continuous parameter interpolating
  between the BH and wormhole solutions describes a wormhole near the ``transition,''
  the quasinormal ringing is represented by damping oscillations appropriate for the near-threshold
  BH solution, but modified by echoes at later times. When the continuous parameter is
  further increased, then the echoes go over into the characteristic quasinormal ringing of the
  wormhole, while the period of the initial ``threshold BH phase'' damped oscillations
  diminishes and looks more like an initial outburst. The term ``transition'' must certainly be understood
  here with a considerable reservation: strictly speaking, we can only talk about values of this
  parameter making the geometry of a wormhole better or worse BH mimicker. Nevertheless,
  if one supposes that this parameter adiabatically changes with time as a result of some
  brane-world dynamic process, then the time-domain profiles of perturbations which relax at a
  much higher rate than the adiabatic change of the metric are qualitatively the same, as was shown,
  for example, in  \cite{Abdalla:2006vb}.

  The paper is organized as follows. In Sec. II we briefly summarize the basic information about the
  brane-world model under consideration. Sec. III discusses the wave equations for test scalar
  and electromagnetic fields and the WKB and time-domain integration methods to be used
  for the analysis of ringing. Sec. IV is devoted to the quasinormal ringing of a few examples of
  BH and wormhole solutions with special emphasis to the parametric range near the
  transition between them. Finally, in the Conclusions we summarize the obtained results.

\section{The brane-world model}

We will consider the second Randall-Sundrum brane-world model (RS2) \cite{ransum2} implying
  that our four-dimensional world is a hypersurface supporting all matter fields and embedded
  in a $\Z_2$-symmetric five-dimensional spacetime (asymptotically AdS bulk), while the
  gravitational field propagates in the whole bulk. The gravitational field on the brane itself
  is described by the modified Einstein equations derived by Shiromizu, Maeda and Sasaki
    \cite{maeda99}
\begin{eqnarray} \label{EE4}
    G\mN = - \Lambda_4\delta\mN -\kappa_4^2 T\mN
	        - \kappa_5^4 \Pi\mN - E\mN,
\end{eqnarray}
    where $G\mN = R\mN - \half \delta\mN R$ is the 4D Einstein tensor,
    $\Lambda_4$ is the 4D cosmological constant expressed in terms of
    the 5D cosmological constant $\Lambda_5$ and the brane tension $\lambda$:
\begin{equation}\label{La4}
    \Lambda_4 = \Half \kappa_5^2
    \biggl(\Lambda_5 + \frac{1}{6} \kappa_5^2\lambda^2\biggr);
\end{equation}
    $\kappa_4^2 = 8\pi G_N = \kappa_5^4 \lambda/(6\pi)$ is the 4D gravitational constant
    ($G_N$ is the Newtonian constant of gravity); $T\mN$ is the stress-energy tensor of matter
    located on the brane; $\Pi\mN$ is a tensor quadratic in $T\mN$, obtained from the
    matching conditions for the 5D metric across the brane:
\beq   \label{Pi_}
    \Pi\mN = \fract{1}{4} T_\mu^\alpha T_\alpha^\nu - \half T T\mN
           - \fract{1}{8} \delta\mN
    \left( T_{\alpha\beta} T^{\alpha\beta} -\thd T^2\right)
\eeq
    where $T = T^\alpha_\alpha$; lastly, $E\mN$ is the so-called ``electric'' part of
    the 5D Weyl tensor projected onto the brane: in proper 5D coordinates, we have
    $E\mn = \delta_\mu^A \delta_\nu^C {}^{(5)} C_{ABCD} n^B n^D$, where the capital letters
    $A, B, \ldots$ are 5D indices, and $n^A$ is the unit normal vector to the brane.
    By construction, $E\mN$ is traceless, that is, $E_\mu^\mu = 0$ \cite{maeda99}.
    The general class and a number of particular examples describing wormholes and
    BHs in the above brane-world scenario were obtained in \cite{bwh1, Bronnikov:2003gx}.

   A feature of utmost interest in these models is the generic appearance of families of
   solutions that unify symmetric wormholes and globally regular BHs with a minimum
   of  the variable $r$ (bounce) in their T-regions and a Kerr-like global structure.
   These two qualitatively different parts of any such family are separated by an
   extremal BH solution.

   In what follows we will consider different examples of such solutions: one family representing
   the generic situation, another one with only an extremal BH, and the third one with a
   zero \Sch\ mass. In each of these examples the metric is a vacuum solutions to \eqs
   \eqref{EE4}, with $T\mN =0$ and $\Lambda_4 =0$; the effective energy-momentum tensor
   in the r.h.s. of \eqref{EE4} is thus the ``tidal'' tensor $E\mN$ of bulk origin. All metrics
   under consideration are \asflat\ and have a zero Ricci scalar.

\section{The methods}

Quasinormal modes of BHs are proper oscillations  of BHs under
  specific boundary conditions, corresponding to purely outgoing waves at infinity and
  purely incoming waves at the event horizon (minus infinity in terms of the tortoise coordinate).
  The boundary conditions for a traversable wormhole which connects two infinities are the same
  in terms of the tortoise coordinate \cite{Konoplya:2005et}, so that many of the tools used for
  finding BH quasinormal modescan, with small modifications, be used for wormholes as well \cite{Aneesh:2018hlp,Konoplya:2016hmd,Konoplya:2010kv,Konoplya:2016hmd,
  Churilova:2019qph,Konoplya:2018ala,Bronnikov:2012ch,Cuyubamba:2018jdl}.
  In particular, to finding the frequencies of quasinormal modes, we will use the WKB method
  when the effective potential has a single maximum and time domain integration
  for all types of the effective potentials.

\subsection{Wave equations}

  The metric of a spherically symmetric static space-time can be written in the general
  form\footnote
  		{Note that we are using notations different from those in \cite{Bronnikov:2003gx}:
  		$B(r)$ of the present paper is equal to $1/B(r)$ of \cite{Bronnikov:2003gx}.}
\beq 		\label{metric}
	ds^2 = -A(r) dt^2+ B(r){dr^2}+r^2 (\sin^2 \theta d\phi^2+d\theta^2).
\eeq
 The generally covariant equation for a massless scalar field has the form
\begin{equation}		\label{KGg}
	\frac{1}{\sqrt{-g}}\d_\mu \left(\sqrt{-g}g^{\mu \nu}\d_\nu\Phi\right)=0
\end{equation}
  and for an electromagnetic field
\begin{equation} \label{EmagEq}
	\frac{1}{\sqrt{-g}}\d_\mu \left(F_{\rho\sigma}g^{\rho \nu}g^{\sigma \mu}\sqrt{-g}\right)=0\,,
\end{equation}
  where $F_{\rho\sigma}=\d_\rho A_{\sigma} - \d_\sigma A_{\rho}$ and $A_\mu$ is the
  vector potential. After separation of variables, Eqs. (\ref{KGg}) and (\ref{EmagEq}) take
  the following general Schr\"odinger-like form:
\begin{equation}\label{wave-equation}
	\frac{d^2\Psi_s}{dr_*^2}+\left(\omega^2-V(r)\right)\Psi_s=0,
\end{equation}
  where $s=0$ corresponds to the scalar field and $s=1$ to the electromagnetic field, the
  ``tortoise coordinate'' $r_*$ is defined by the relation
\begin{equation}
	dr_*=dr\sqrt{\frac{B(r)}{A(r)}},
\end{equation}
  and the effective potentials are (see, e.g.,~\eq(12a) in \cite{Zinhailo:2018ska})
\bearr           \label{spotential}
	V_{s}(r) = A(r) \frac{\ell(\ell+1)}{r^2}+\frac{1}{2r}\frac{d}{dr}\frac{A(r)}{B(r)},
\yyy                      \label{empotential}
	V_{em}(r) = A(r)\frac{\ell(\ell+1)}{r^2}.
\ear

\subsection{The WKB approach}

 For the analysis in the frequency domain we shall use the semi-analytical WKB method
  \cite{Schutz:1985zz,Iyer:1986np,Matyjasek:2017psv,Konoplya:2003ii,Konoplya:2019hlu}.
  The essence of this approach is the expansion of the solution at both infinities in WKB series
  and matching these asymptotic expansions with the Taylor expansion near the peak of the
  effective potential. In addition, according to \cite{Matyjasek:2017psv}, we use a further
  representation of the WKB expansion in the form of the Pad\'e approximants which, in most
  cases, greatly improves the accuracy of the WKB method. The WKB formula can be written
  in the following form \cite{Konoplya:2019hlu}:
\bearr               \label{WKBformula-spherical}
	\omega^2 = V_0+A_2(\K^2)+A_4(\K^2)+A_6(\K^2)+\ldots
\nnn
	- \imo \K\sqrt{-2V_2}\left(1+A_3(\K^2)+A_5(\K^2)+A_7(\K^2)\ldots\right),
\nnn	
\ear
  where $\K=n+1/2$, $n=0,1,2,3\ldots$.

  The corrections $A_k(\K^2)$ of order $k$ to the eikonal formula are polynomials in $\K^2$
  with rational coefficients and depend on the values $V_2, V_3\ldots$ of higher-order derivatives
  of the potential $V(r)$ at its maximum. To increase accuracy of the WKB formula, we use the
  procedure suggested by Matyjasek and Opala \cite{Matyjasek:2017psv}, which consists in
  using the Pad\'e approximants. For the order $k$ of the WKB formula
  (\ref{WKBformula-spherical}) we define a polynomial $P_k(\epsilon)$ in the following way
\bearr
	P_k(\epsilon)=V_0+A_2(\K^2)\epsilon^2+A_4(\K^2)\epsilon^4+A_6(\K^2)\epsilon^6+\ldots
\nnn	
	-\imo \K\sqrt{-2V_2}\left(\epsilon+A_3(\K^2)\epsilon^3+A_5(\K^2)\epsilon^5\ldots\right),
		\label{WKBpoly}
\ear
  and the squared frequency is obtained for $\epsilon=1$:
\[
		\omega^2=P_k(1).
\]

  For the polynomial $P_k(\epsilon)$ we will use Pad\'e approximants
\begin{equation}\label{WKBPade}
	P_{\tilde{n}/\tilde{m}}(\epsilon)=\frac{Q_0+Q_1\epsilon+\ldots+Q_{\tilde{n}}\epsilon^{\tilde{n}}}		
			{R_0+R_1\epsilon+\ldots+R_{\tilde{m}}\epsilon^{\tilde{m}}},
\end{equation}
  with $\tilde{n}+\tilde{m}=k$, such that, near $\epsilon=0$,
\[
	P_{\tilde{n}/\tilde{m}}(\epsilon)-P_k(\epsilon)={\cal O}\left(\epsilon^{k+1}\right).
\]

  Usually, for finding the fundamental mode ($n=0$) frequency, Pad\'e approximants with
  $\tilde{n}\approx\tilde{m}$ provide the best approximation. In \cite{Matyjasek:2017psv},
  $P_{6/6}(1)$ and $P_{6/7}(1)$ were compared to the 6th-order WKB formula $P_{6/0}(1)$.
  In \cite{Konoplya:2019hlu} it has been observed that usually even $P_{3/3}(1)$, i.~e. a
  Pad\'e approximation of the 6th-order gives a more accurate value for the squared frequency
  than $P_{6/0}(1)$. Here we will use the 7th WKB expansions with $\tilde{m}=3$ Pad\'e
  approximation and show that the results obtained at different WKB orders are in a very
  good agreement.

\subsection{Time-domain integration}

  We will integrate the wavelike equations rewritten in terms of the light-cone variables $u=t-r_*$
  and $v=t+r_*$. The appropriate discretization scheme was suggested in \cite{Gundlach:1993tp}:
\bearr                       \label{Discretization}
	\Psi(N) = \Psi(W)+\Psi(E)-\Psi(S)
\nnn \	
	- \Delta^2\frac{V(W)\Psi(W)+V(E)\Psi(E)}{8}+{\cal O}(\Delta^4)\,,
\ear
  where we use the following designations for the points:
  $N=(u+\Delta,v+\Delta)$, $W=(u+\Delta,v)$, $E=(u,v+\Delta)$ and $S=(u,v)$. The initial data
  are specified on the null surfaces $u=u_0$ and $v=v_0$.  To extract the values of the
  quasinormal frequencies, we will use the Prony method which allows us to fit the signal by
  a sum of exponentials with some excitation factors.

\section{The picture of quasinormal ringing for various solutions}

\begin{figure*}
\resizebox{\linewidth}{!}{\includegraphics*{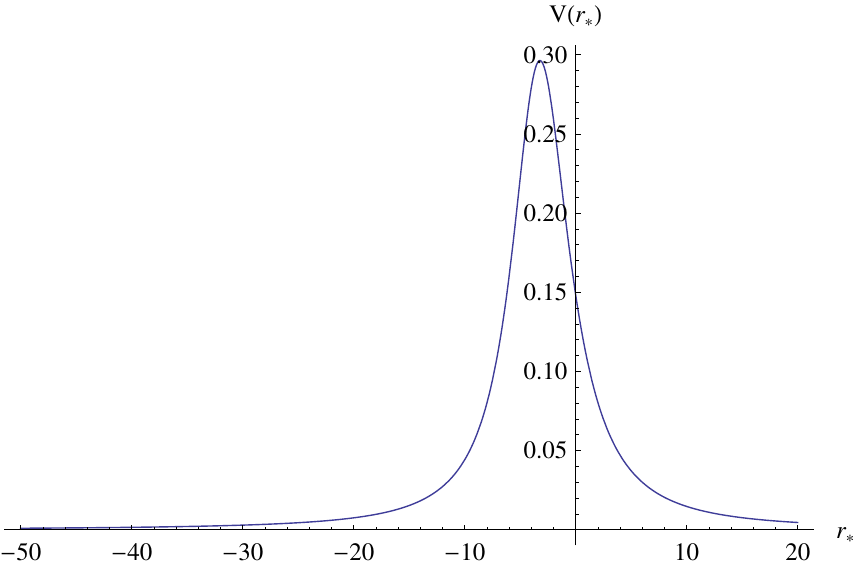}\includegraphics*{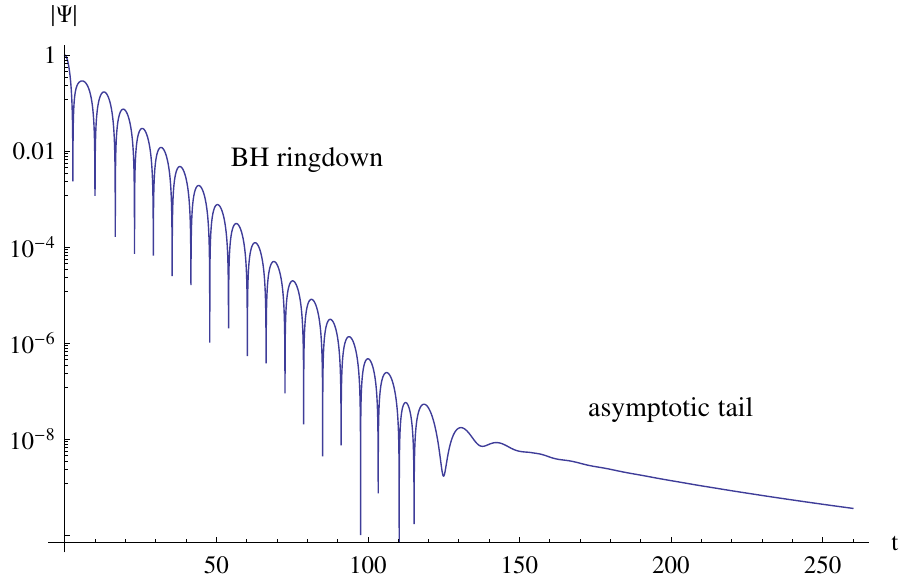}}
\caption{The effective potential (left) and the semi-logarithmic plot of  the time-domain profile (right)
	for perturbations of the electromagnetic field in the vicinity of the double horizon of
	CFM BHs with the metric \ref{ds-x1} where $r_{0} = 2 M$, $M=1/2$, $\ell=1$. The
	quasinormal ringing goes over into asymptotic power-law tails.}\label{fig0}
\end{figure*}
\begin{figure*}
\resizebox{\linewidth}{!}{\includegraphics*{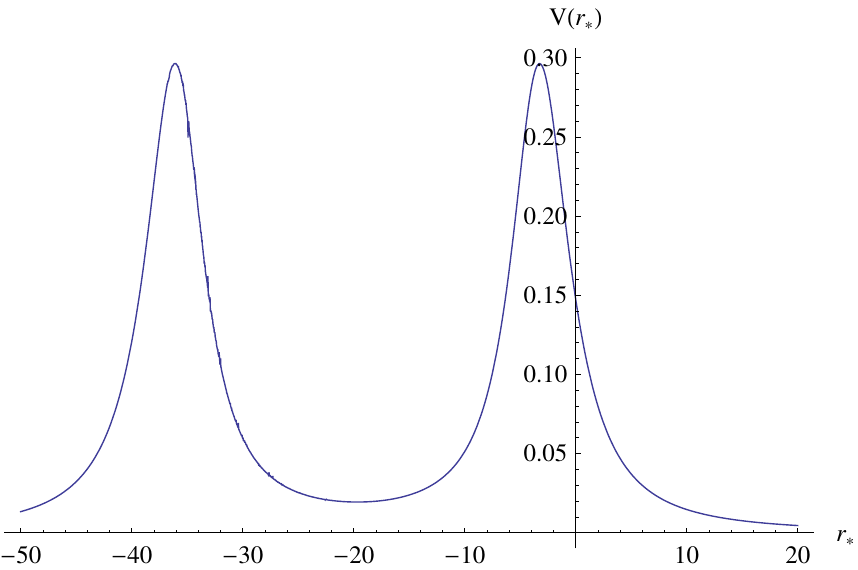}\includegraphics*{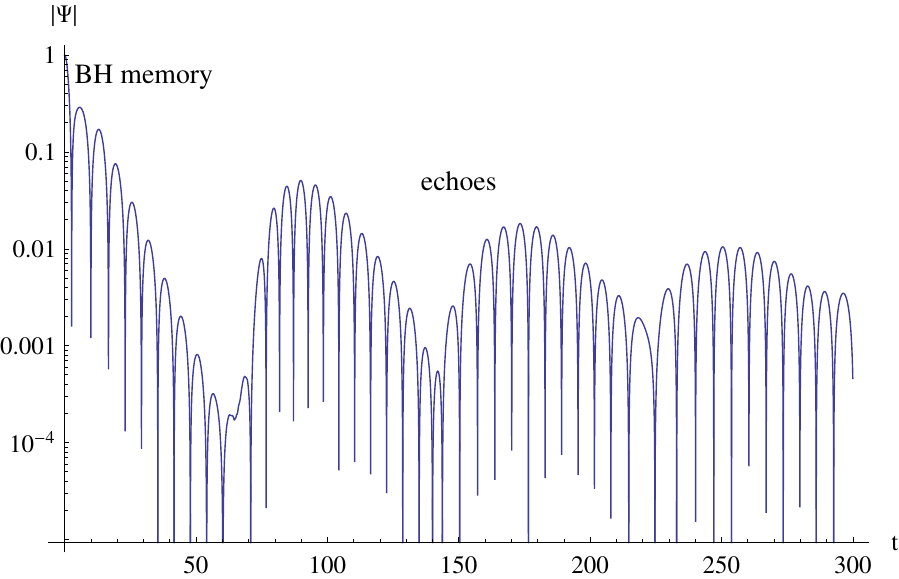}}
\caption{The effective potential (left) and the semi-logarithmic plot of  the time-domain profile (right)
	for perturbations of the electromagnetic field in the vicinity of a CFM wormhole
	with $M=1/2$, $r_{0} =1.01$, $\ell=1$. The first long period of damped oscillations is
	indistinguishable from that of the threshold BH metric (that corresponds to $r_{0} = 2 M$),
	but, at later times it goes over into a series of echoes.}\label{fig1}
\end{figure*}
\begin{figure*}
\resizebox{\linewidth}{!}{\includegraphics*{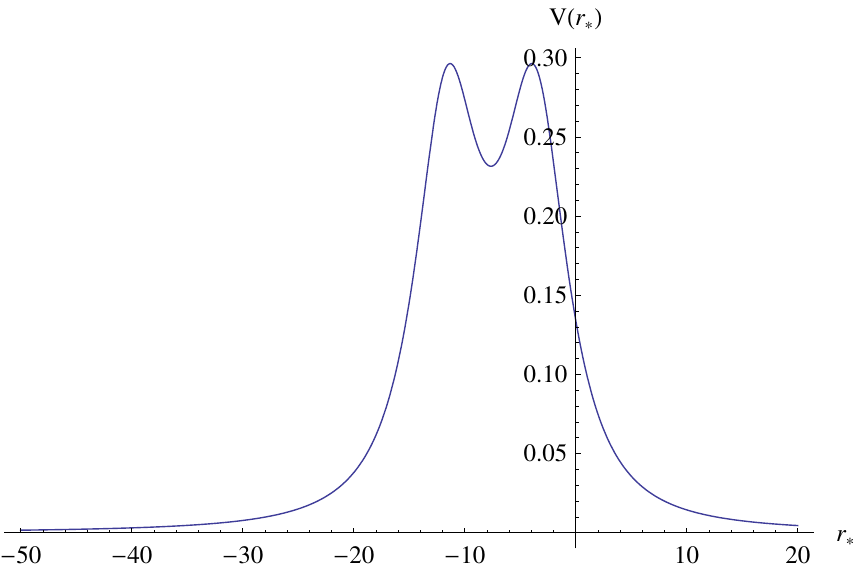}\includegraphics*{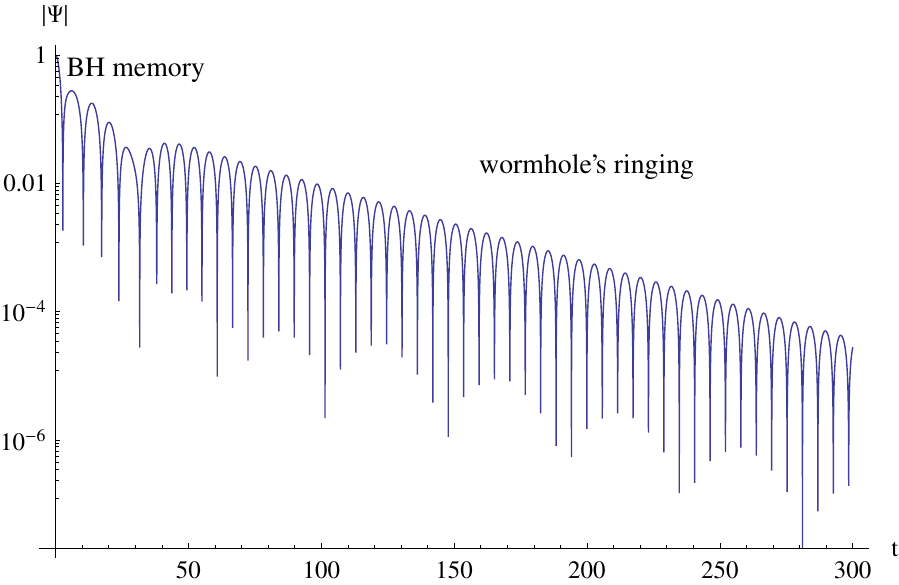}}
\caption{The effective potential (left) and the semi-logarithmic plot of  the time-domain profile (right)
	for perturbations of the electromagnetic field in the vicinity of a CFM wormhole
	with $M=1/2$, $r_{0} =1.2$, $\ell=1$. The first relatively short period of damped oscillations
	is a kind of memory of the threshold BH state $r_{0} = 2 M$, and it goes over into
	the characteristic quasinormal mode of the wormhole at later times.}\label{fig2}
\end{figure*}
\subsection{The Casadio-Fabbri-Mazzacurati (CFM) metric}
Our first example with the \Sch\ generating function $A(r) = 1 - 2M/r$
   \cite{Bronnikov:2003gx} represnts the generic behavior of brane-world BH/wormhole families.
  The metric has the form
\bearr \label{ds-x1}
	ds^2 = \schd dt^2
\nnn \qquad
	     - \frac{1 - 3 M/(2r)}{(1 - 2 M/r)(1 - r_0/r)}dr^2 -r^2 d\Omega^2.
\ear
    The Schwarzschild metric is reproduced in the special case $r_0 = 3 M/2$. The
    metric (\ref{ds-x1}) has been obtained by Casadio, Fabbri and Mazzacurati
    \cite{casad01} in search for new brane-world BH solutions and by Germani and
    Maartens \cite{maart01} as a possible  metric outside a homogeneous star on the brane.

    In the case $r_0 > 2 M$, the metric (\ref{ds-x1}) describes a symmetric traversable
    \wh\ \cite{bwh1}.

    In the intermediate case $r_0 = 2 M$ we obtain a BH with a double horizon at $r = 2 M$.

    In the case $r_0 < 2 M$, there is a BH with a single horizon at $r=2 M$. As described in
    Ref.\,\cite{casad01}, the global space-time structure depends on the sign of
    $\eta = r_0 - 3 M/2$. If $\eta < 0$, this structure coincides with that of a Schwarzschild BH,
    but the spacelike curvature singularity is located at $r= 3 M/2$ instead of $r=0$. If
    $\eta > 0$, the solution describes a nonsingular BH with a minimum value of
    the variable $r$ equal to $r_0 > 3M/2$  inside the horizon, that is, a bounce in the two
    angular directions of a Kantowski-Sachs anisotropic cosmology.

\begin{table*}
\centering
\caption{Quasinormal mode frequencies $\omega$ of the electromagnetic field
	for $\ell=1$ in the CFM metric.\ref{ds-x1}}\label{tab1}
\medskip	
\begin{tabular}{p{1cm}cccc}
\hline
$r_{0}$ & Time-domain  & WKB \\
\hline
0      & $0.43751 - 0.24617 i$ & $0.43782-0.24904 i$  \\
0.2    & $0.45313 - 0.23549 i$ & $0.45365-0.23682 i$  \\
0.4    & $0.46936 - 0.22148 i$ & $0.46942-0.22174 i$  \\
0.6    & $0.48557 - 0.20288 i$ & $0.48509-0.20306 i$  \\
0.75   & $0.49653 - 0.18495 i$ & $0.49652-0.18498 i$  \\
0.8    & $0.49983 - 0.17789 i$ & $0.50056-0.17729 i$  \\
1      & $0.50651 - 0.14686 i$ & $0.50652-0.14733 i$  \\
1.1    & $0.50177 - 0.11188 i$, echoes & --  \\
1.2    & $0.48143 - 0.06821 i, 0.99460 - 0.00979 i$ & -- \\
1.3    & $0.47713 - 0.07202 i, 0.56376 - 0.04650 i$ & --  \\
10    & initial outburst, $0.12759 - 0.03054 i $ & --  \\
20    & initial outburst, $0.06552 - 0.01498 i $ & --  \\
\hline
\end{tabular}
\end{table*}

\begin{table*}
\centering
\caption{Quasinormal modes of the electromagnetic field for $\ell=1$ for the BK-1 metric
	\ref{f-x3}.}\label{tab1}
\medskip
\begin{tabular}{p{1cm}cccc}
\hline
$r_{0}$ & Time-domain  & WKB \\
\hline
1      & $0.53569 - 0.06300 i$ & $0.53569 - 0.06299 i$  \\
1.1    & $0.53360 - 0.06227 i$, echoes & --  \\
1.2    & $0.51232 - 0.05364 i$, echoes & --  \\
1.3    & $0.50762 - 0.04834 i$, echoes & --  \\
1.4    & $0.49793 - 0.04302 i, 0.55835 - 0.01965 i$ & -- \\
1.5    & $0.47806 - 0.03107 i, 0.55859 - 0.02937 i$ & -- \\
\hline
\end{tabular}
\end{table*}
\begin{figure*}
\resizebox{\linewidth}{!}{\includegraphics*{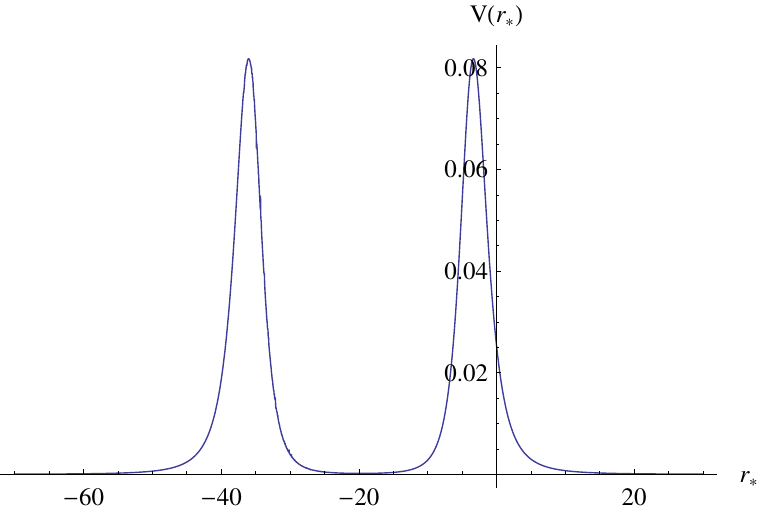}\includegraphics*{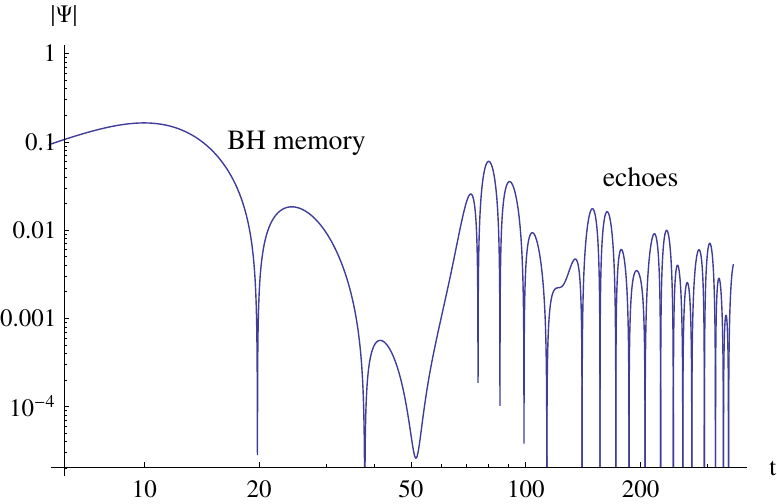}}
\caption{The effective potential (left) and the logarithmic plot of the time-domain profile (right)
	for perturbations of the scalar field in the vicinity of the CFM wormhole with
	$M=1/2$, $r_{0} =1.01$, $\ell=0$. Several first oscillations (usual for $\ell=0$ perturbations)
	are indistinguishable from those of the threshold BH metric (corresponding to $r_{0} = 2 M$),
	but at later times they go over into a series of echoes.}\label{fig4}
\end{figure*}

  In Fig.\,\ref{fig0} one can see the usual picture of evolution of perturbations in the vicinity of
  a BH, now for the threshold case $r_{0}=2 M$. It consists of the initial outburst which
  changes into damped quasinormal oscillations and power-low asymptotic tails in the end.
  Figure\,\ref{fig1} shows perturbations of a wormhole corresponding to $r_{0} = 2.02 M$, that is,
  very close to the threshold with the BH state, and there emerges a distinctive picture of
  echoes after the initial oscillatory fall-off corresponding to the residual fundamental mode of
  the boundary $r_{0}=2 M$ BH. Finally, when the parameter $r_{0}$ is further increased, the
  echoes go over into the established characteristic quasinormal ringing of a wormhole
  (Fig.\,\ref{fig2}).

 Table I exhibits the dominant quasinormal frequencies of the CFM metric in both regimes.
  In the BH case $r_{0}\leq 2 M$, the quasinormal modes can be computed with both
  time-domain integration and WKB methods. Near the threshold, when $r_{0} \gtrapprox 2 M$,
  the WKB formula, implying two turning points, cannot be used since the effective potential has
  two barriers and four turning points.

  We can see that for BHs the results produced by the WKB method and time-domain integration
  are in a very good concordance, with disagreement always less than one percent. Taking
  into account that the extraction of frequencies from the profiles obtained with the help of
  time-domain integration greatly depends on the temporal range which is determined as the
  ``quasinormal ringing,'' a relative error within one percent can be easily ascribed not even to
  the WKB formula but rather to the arbitrariness of the period of quasinormal ringing. Here we
  used the 7th order WKB formula with the further Pad\'e approximants such that $\tilde{m}=3$,
  which gives the best agreement with the results of time-domain integration for the
  Schwarzschild case $r_{0} = 3/(2 M)$ and also coincides (within a 5-digits accuracy) with
  the accurate numerical value $0.49652-0.18498 i$. For the case of a wormhole very close
  to the threshold, we have only a single mode which is the residual or, in a sense, ``memory'' of
  the boundary BH solution. The echoes which follow this residual mode cannot be represented
  by a single dominant quasinormal mode. Nevertheless, when the parameter $r_{0}$ is
  further increased, the enveloping oscillations of echoes (visible in Fig.\,\ref{fig1}) align, and
  the characteristic mode of the wormhole establishes, being reflected in the second frequency
  given in Table 1. We can also see that at $r_{0}\gg 2 M$ the real and imaginary parts
  of the frequency are inversely proportional to $r_{0}$.

The above description concerned the behavior of electromagnetic perturbations against the
  background od brane-world BHs and \whs. It can be verified  that massless scalar field
  perturbations show similar qualitative features, as can be seen in Fig.\,4 that depicts the
  behavior of radial scalar perturbations ($\ell =0$) of a CFM \wh.

\subsection{The Bronnikov-Kim-1 (BK-1) metric}

This case represents an atypical example of a BH-\wh\ family of solutions.
   The metric can be written as
\bearr       \nhq           \label{f-x3}
	  ds^2 = \schd^2 dt^2 - \biggl(1-\frac{r_0}{r}\biggr)^{-1}
	        \biggl(1-\frac{r_1}{r}\biggr)^{-1} dr^2
\nnn \cm
		-  r^2 d\Omega^2, \cm             r_1 := \frac{M r_0}{r_0- M},
\ear
  The only BH solution corresponds to the case $r_0 = r_1 = 2 M$, which coincides with the
  extremal Reissner-Nordstro\"m metric.

  Other values of $r_0$ lead either to \whs\ (the throat is located at $r = r_0$ if $r_0 > 2 M$
  or at $r = r_1 > 2 M$ in case $2 M > r_0 > M$), or to a naked singularity located at
  $r=2 M$ (if $r_0 < M$) (see more details in Ref.\,\cite{bwh1}).

\begin{figure*}
\resizebox{\linewidth}{!}{\includegraphics*{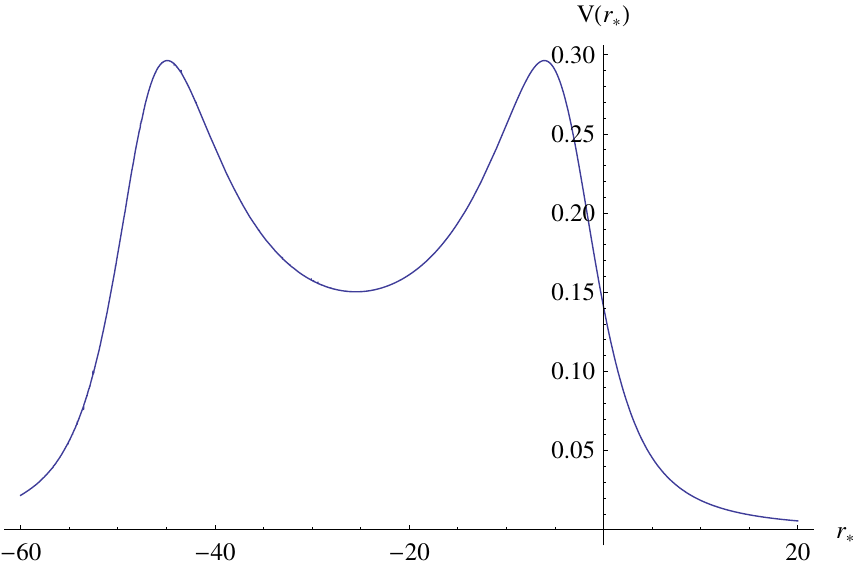}\includegraphics*{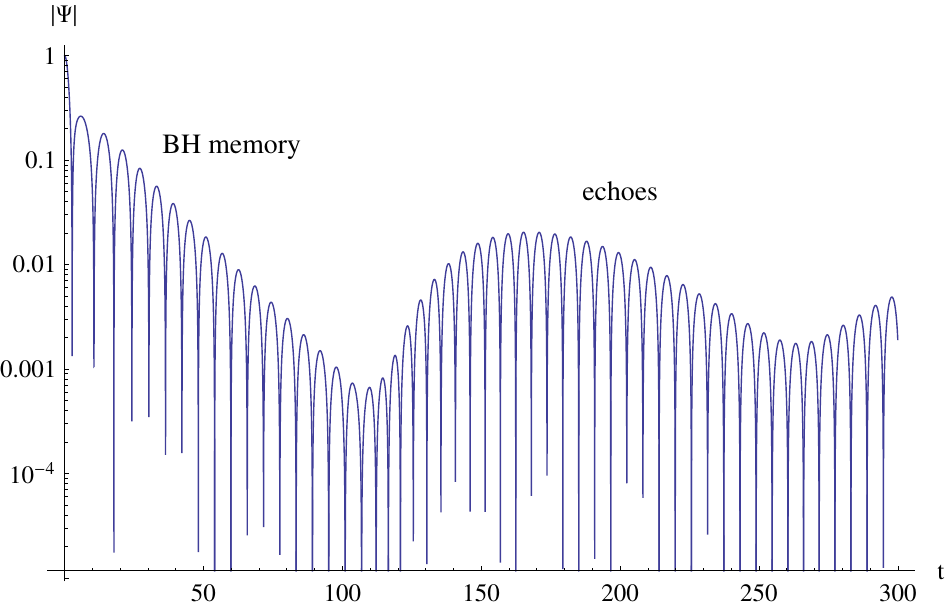}}
\caption{The effective potential (left) and the semi-logarithmic plot of the time-domain profile
	(right) for perturbations of the electromagnetic field in the vicinity of the BK-1 wormhole:
	$M=1/2$, $r_{0} =1.1$, $\ell=1$. The first long period of damped oscillations is
	indistinguishable from that of the threshold BH metric (that corresponds to $r_{0} = 2 M$),
	but at later times emerges a series of echoes.}\label{fig3}
\end{figure*}

\begin{figure*}
\resizebox{\linewidth}{!}{\includegraphics*{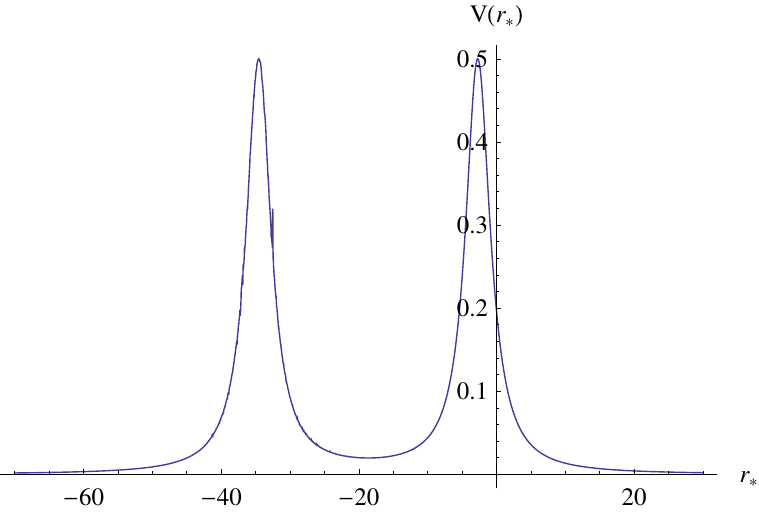}\includegraphics*{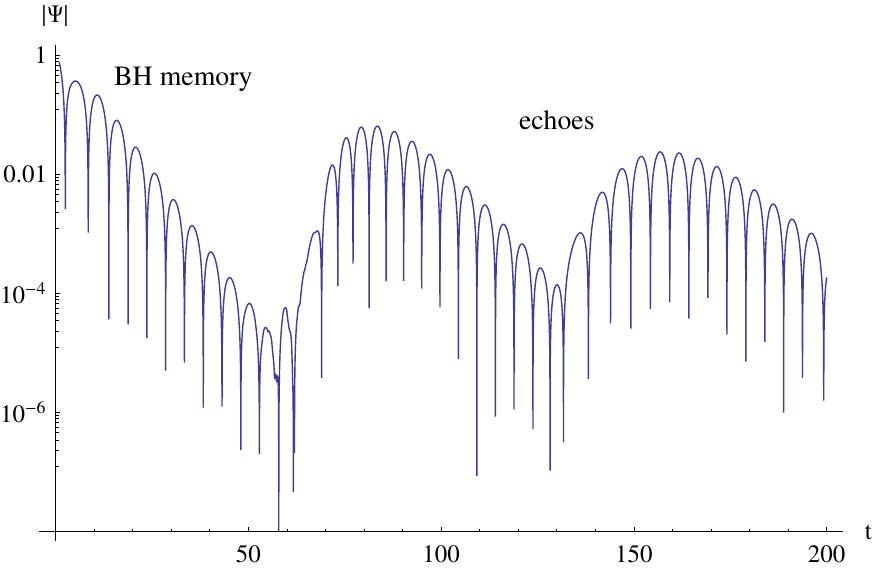}}
\caption{
The effective potential (left) and the semi-logarithmic plot of  the time-domain profile
	(right) for perturbations of the electromagnetic field in the vicinity of the BK-2 wormhole
	\ref{f-x2} with $M=1/2$, $C =-0.01$, $\ell=1$. The first long period of damped oscillations
	is indistinguishable from that of the threshold BH metric (corresponding to $C = 0$), but
	at later times it goes over into a series of echoes.
}\label{fig6}
\end{figure*}
\begin{table*}
\centering
\caption{Quasinormal modes of the electromagnetic field for $\ell=1$ in the BK-2 metric \ref{f-x2}.
}\label{tab1}
\medskip
\begin{tabular}{p{1cm}cccc}
\hline
$C$ & Time-domain  & WKB \\
\hline
1.5      & $ 0.53695 -0.34301 i$ & $0.54096-0.33905 i$  \\
1    & $ 0.57668 -0.31754 i$ & $0.57663-0.31785 i$  \\
0.5    & $ 0.61726 - 0.27547 i$ & $0.61983-0.27457 i$  \\
0.1    & $ 0.64156-0.22399 i$ & $ 0.64144-0.22379 i$  \\
0   & $0.64439 - 0.20921 i$ & $0.64435-0.20900 i$  \\
-0.01  & $0.64306 - 0.20529 i$, echoes & --  \\
-0.1   & $0.61509 - 0.09685 i$, echoes & --  \\
-0.3   & $0.59897 - 0.07043 i$, $0.70323 -0.03991 i$ & --  \\
-0.5     & initial outburst, $0.73833 -0.07600 i$ & --  \\
-0.7     & initial outburst, $ 0.73338 - 0.09961 i  $ & --  \\
\hline
\end{tabular}
\end{table*}

  Figure\,\ref{fig3} shows the picture of evolution of perturbation in the vicinity of the threshold,
  which is qualitatively the same as for the previous metric, with a distinction related to the
  length of the initial (residual) quasinormal ringing which is much longer now, and also the
  frequencies change slower with changing $r_{0}$.  This picture simply reflects the fact that
  larger $r_{0}$ are necessary to deviate from the threshold BH geometry $r_{0}= 2 M$
  at the same extent.

\subsection{The Bronnikov-Kim-2 (BK-2) metric}

This example differs from the previous ones by having a zero value of the \Sch\ mass
	of the brane-world BHs and \whs. The metric is
\bearr
	ds^2 = \biggl(1-\frac{h^2}{r^2}\biggr) dt^2
\nnn      \label{f-x2}
	- \biggl(1-\frac{h^2}{r^2}\biggr)^{-1} \biggl(1 + \frac{C-h}{\sqrt{2r^2 - h^2}}\biggr)^{-1} dr^2
	 -  r^2 d\Omega^2
\ear
  The sphere $r=h$ is a simple horizon if $C >0$ and a double horizon if $C=0$. In the case
  $C < 0$, $B(r)$ has a simple zero at $r =r_{\rm th} > h$ given by
\beq
	    2 r^2_{\rm th} = h^2 + (h - C)^2,                       \label{th-x2}
\eeq
    which is a symmetric \wh\ throat \cite{bwh1}.

  The evolution of perturbations in this example is qualitatively similar to the two previous cases:
  the echoes appears immediately after the threshold, providing wormholes mimicking the BH
  behavior. If we consider the quality factor of the mode, which is proportional to the ratio of the
  real oscillations frequency to the damping rate,
\[
	Q \sim \frac{\re{\omega}}{|\im {\omega}|},
\]	
  we can see that oscillations of BHs have a maximum quality factor in the extremal state
  at the threshold. This is true not only for the last example but for all models under consideration.
  Moreover, if one considers wormholes with the parameter $C$ (or $r_{0}$ in the previous
  examples) near the threshold, one can see that the quality factor continues growing and
  starts to decrease only when echoes are damped. This is also a common feature of all
  three examples.

\section{Conclusions}

  We have used the higher-order WKB method with Pad\'e approximants  \cite{Schutz:1985zz,Iyer:1986np,Konoplya:2003ii,Matyjasek:2017psv,Konoplya:2019hlu}
  and time-domain integration \cite{Gundlach:1993tp} in order to analyze the quasinormal
  ringing of BH and wormhole solutions in the RS2 brane-world model. The metrics under
  consideration depend on a continuous parameter interpolating between BHs and wormholes,
  so that if the parameter is larger than some threshold value, then the BH ``goes over'' into
  a wormhole. We have shown that this ``transition'' from BHs to wormholes near the threshold
  is characterized by echoes: the first stage of damped oscillations, representing the memory
  of the threshold BH state is accompanied by a series of echoes at later times. When the same
  parameter is further increased, the echoes damp and pass on into the characteristic ringing of
  the wormhole.

  This picture is observed for all examples considered here and, apparently, does not depend
  on the particular model. We believe that this kind of behavior should not depend on the spin
  of the field as well and should be valid not only for the scalar and electromagnetic fields
  considered here, but also for the gravitational field because the echoes are induced by
  appearing of the second symmetric peak in the far left region (in terms of the tortoise coordinate),
  producing the second scattering and partial reflection of the signal at late times.

  Although the characteristic dominant quasinormal frequency of the BHs and wormholes
  under consideration can behave differently in different examples, there is one feature observed
  in all the considered models. The ratio of the real oscillation frequency to the damping rate,
  which is proportional to the quality factor of oscillations, always decreases as the BH
  solution moves away from the threshold. On the other hand, the quality  factor continues
  increasing on the other side of the threshold for wormholes which still mimic the BH behavior,
  that is, for the residual mode. In other words, wormholes near the threshold are the best oscillators
  among all the considered examples. Therefore, it is not excluded that this monotonic behavior
  of the quasinormal frequency might be a more general property of the brane-world models.

\acknowledgments{
The authors acknowledge  the  support  of  the  grant  19-03950S of Czech Science Foundation ($GA\check{C}R$). The work of K.B. was partly performed within the framework of the Center FRPP supported by
MEPhI Academic Excellence Project (contract No. 02.a03.21.0005, 27.08.2013).

The work was also funded by the RUDN University Program 5-100 and by the RFBR grant No. 19-02-00346.
}


\end{document}